\documentclass[aps,prl,floats,floatfix,twocolumn,superscriptaddress,10pt,nofootinbib]{revtex4-2}

\usepackage{latexsym}
\usepackage{amsmath,amssymb,amsthm,thmtools}
\usepackage{mathtools,mathrsfs}
\usepackage{amsbsy}
\usepackage{soul}
\usepackage[caption=false]{subfig}
\usepackage{bbold}
\usepackage{comment}

\usepackage[utf8]{inputenc}

\usepackage[pdftex]{graphicx}
\graphicspath{{./figures/}}

\usepackage[usenames,dvipsnames,table]{xcolor}
\usepackage{float}

\usepackage[colorlinks=true]{hyperref}
\usepackage[nameinlink]{cleveref}
\hypersetup{
  colorlinks   = true, 
  urlcolor     = green!80!black, 
  linkcolor    = blue, 
  citecolor    = red!80!black 
}

\Crefname{figure}{Figure}{Figures}
\crefname{figure}{Fig.}{Figs.}

\usepackage{graphicx}

\usepackage[normalem]{ulem}

\usepackage{microtype}

\usepackage{physics}

\newcommand{\on}[1]{{\operatorname{#1}}}

\newcommand{\calE}{\mathcal{E}}
\newcommand{\calF}{\mathcal{F}}
\newcommand{\calI}{\mathcal{I}}

\newcommand{\calS}{\mathcal{S}}

\newcommand{\Red}{\mathrm{Red}}

\begin{document}

\title{
On the emergence of quantum Darwinism and pointer states for non-commuting evolutions 
}

\author{Diana A.~Chisholm}
\affiliation{Universit\`a degli Studi di Palermo, Dipartimento di Fisica e Chimica -- Emilio Segr\`e,\\ via Archirafi 36, I-90123 Palermo, Italy}
\author{G.~Massimo Palma}
\affiliation{Universit\`a degli Studi di Palermo, Dipartimento di Fisica e Chimica -- Emilio Segr\`e,\\ via Archirafi 36, I-90123 Palermo, Italy}
\author{Luca Innocenti}
\affiliation{Universit\`a degli Studi di Palermo, Dipartimento di Fisica e Chimica -- Emilio Segr\`e,\\ via Archirafi 36, I-90123 Palermo, Italy}

\begin{abstract}    
    Quantum systems achieve objectivity by redundantly encoding information about themselves into the surrounding environment, through a mechanism known as quantum Darwinism. When this happens, observes measure the environment and infer the system to be in one of its pointer states. 
    We study the emergence of objectivity whenever the Hamiltonian of the system and the interaction Hamiltonian between system and environment do not commute, a condition which is thought to be incompatible with the presence of pointer states. We show that, not only non-commuting evolutions allow for the emergence of objective states, but it is possible to give a more relaxed definition of pointer states, that is always well defined whenever there is redundancy of information, and coincides with the usual one for commuting evolutions.
\end{abstract}

\maketitle

\paragraph{Introduction}
In the context of quantum Darwinism~\cite{ollivier_objective_2004,zurek_quantum_2009,zurek2022QuantumTheoryClassical,zurek_quantum_2009, horodecki_quantum_2015,zurek_decoherence_2003,korbicz_roads_2021}, a system is said to be \textit{objective} if many observers can independently infer its state without perturbing it.
Let $\rho_{\calS\calF}$ be the joint state of a system $\calS$ and an environment $\calE=\bigotimes_i \calF_i$ partitioned into fractions $\calF_i$.
We say that $\calS$ exhibits objectivity when information about $\calS$ is redundantly encoded in many environment fractions, that is, $I_{\operatorname{acc}}(\calS:\calF_i)\simeq S(\rho_\calS)$ for all $i$~\cite{le_strong_2019,korbicz_roads_2021}, where $S(\cdot)$ is the von Neumann entropy and $I_{\operatorname{acc}}$ the accessible mutual information~\cite{preskill2016quantum}.
Most models assume unitary evolution of the joint state under a (often time-independent) Hamiltonian in which the system Hamiltonian $H_\calS$ commutes with the system-environment interaction $H_\calI$~\cite{megier_correlations_2022, mirkin_many-body_2021, mironowicz2022non, guo2023relation, engineer2024equilibration}, and all proof-of-principle demonstrations operate within this framework~\cite{ciampini_experimental_2018, chen_emergence_2019, unden_revealing_2019,zhu2025observation,chisholm_witnessing_2021}.
When a dynamic creates objectivity in $\rho_{\calS\calE}$, it has been shown that information about at most one system observable can proliferate and be accessible from the environment~\cite{zurek_pointer_1981,zurek1982environment}; the eigenstates of this observable are the \textit{pointer states}, and can be characterised as the states left pure and unperturbed by the evolution, as the eigenbasis of $H_\calS$, or alternatively as the system basis appearing in the spectrum-broadcast-structure (SBS) decomposition of $\rho_{\calS\calE}$~\cite{korbicz_objectivity_2014,horodecki_quantum_2015,korbicz_roads_2021}.
These different notions are equivalent, as long as $[H_\calS,H_\calI]=0$ holds.

However, the condition $[H_\calS,H_\calI]=0$ is quite restrictive and not justified in many real-world settings.
In particular, it is to be expected that real measurement devices will involve much more complex dynamics and, in particular, non-commuting Hamiltonians.
Relaxing the assumption of commutativity has important consequences on the definition and meaning of pointer states: the above definitions of pointer states no longer coincide, and it becomes unclear whether --- and how --- objectivity and pointer states persist.
Here, we investigate dynamics in the presence of non-commutativity, so that $[H_\calS,H_\calI]\neq0$, and study the emergence of objectivity and pointer states in such settings.
We show that non-commutativity does not preclude objectivity, and that SBS states can still form, enabling us to define \textit{SBS-pointer-states} as the system states whose information is proliferated into the environment.
More specifically, using simple qubit models we show that (1) non-commutativity hinders --- but does not prevent --- objectivity; (2)
the resulting pointer states depend non-trivially on the interplay between $H_\calS$ and $H_\calI$, rather than being defined solely by $H_\calS$;
and (3) the degrading effects of non-commutativity are minimised whenever the interaction dynamic is much faster than the intrinsic system dynamics, a condition that emerges naturally for macroscopic environments~\cite{kicinski2021decoherence, lee2024encoding}, even though $[H_\calS,H_\calI]=0$ does not generally hold in these scenarios.
Our results thus extend the framework of quantum objectivity to the non-commuting regime, and offer valuable insight on the physical meaning of pointer states therein.

\paragraph{The model}
\label{sec:model}

We consider a single-qubit system $\calS$ coupled to $n$ environment qubits $\calE_i$ through the standard dephasing-type interaction  $H_\mathcal{I}=\sum_{i=1}^n H_{\calS\calE_i}$, where each
$H_{\calS\calE_i}=\gamma\sigma_{z}^{\calS}\otimes\sigma_{x}^{\calE_i}$ describes the interaction between $\calS$ and a single environment fraction $\calE_i$~\cite{campbell2019collisional, lorenzo_anti-zeno-based_2020, garcia-perez_decoherence_2020, Chisholm_Stochastic_2021}. 
With no loss of generality for our purposes, we will neglect  the environment free evolution by assuming that its Hamiltonian commutes with the interaction Hamiltonian and thus moving to the interaction picture.
The total Hamiltonian is
\begin{equation}\label{eq1}
    H_{\mathrm{tot}}=H_\calS+H_{\mathcal{I}}.
\end{equation}
To study non-commutativity, we choose $H_\calS=\omega(p\sigma^\calS_{x}+(1-p)\sigma^\calS_{z})$, $p\in[0,1]$,
and use the parameter $p$ to control the degree of non-commutativity, on account of $[H_\calS,H_\calI]\propto p$.
Environmental qubits are initialised in $\ket{0}$, and the system in the $\sigma_{y}$ eigenstate $\ket{\circlearrowleft}=\frac{1}{\sqrt{2}}\left(\ket{0}+i\ket{1}\right)$.
By choosing an initial state that lies in an equal-weight superposition of $|0\rangle$ and $|1\rangle$ one maximizes susceptibility to decoherence induced by $H_{\mathcal I}$, and among among those the eigenstates of $\sigma_y$ are the most sensitive to the \(\sigma_x^\mathcal S\) term in \(H_{\mathcal S}\), thereby making the effects of non-commutativity more prominent.
Different choices of initial states yield qualitatively similar results.
The interaction $H_\calI$ determines the basis in which the environment ``reads'' the system: each environmental qubit $\calE_i$ evolves conditionally under either the $\gamma \sigma_x^{\calE_i}$ or the $-\gamma \sigma_x^{\calE_i}$ Hamiltonian term, undergoing different unitary evolutions conditionally on the system being in the $|0\rangle$ or the $|1\rangle$ state, thus transferring information about the system's $\sigma_z$ eigenvalue into the $\ket\pm$ basis of the environment qubits.
Meanwhile, if $[H_\calS, H_\calI]\neq0$, the populations of the system in the $\sigma_z^\calS$ basis are no longer constants of motion,
so that the information encoded into the environment will progressively become outdated as time advances.

\begin{figure}[t]
    \centering
    \includegraphics[width=\columnwidth]{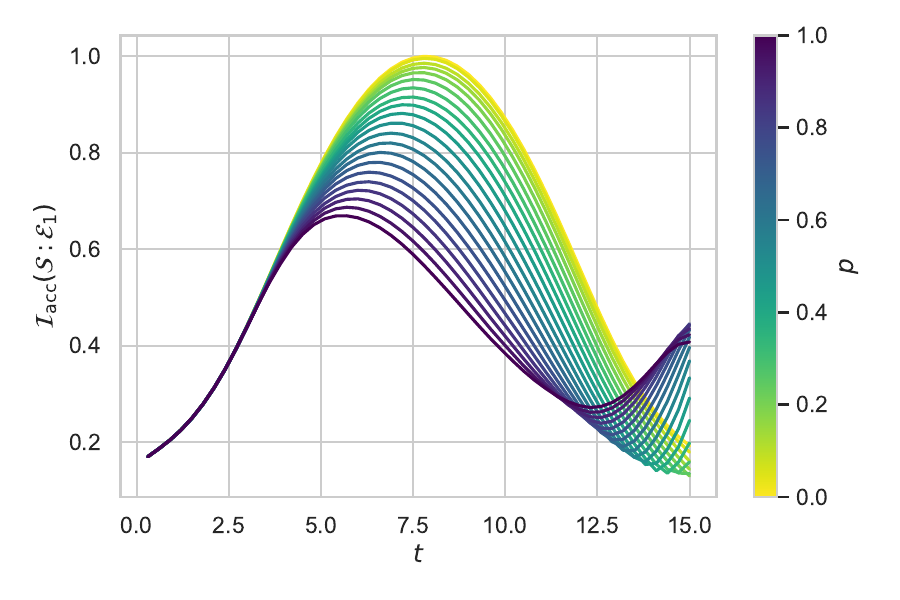}
    \caption{
    \textbf{Accessible information as a function of time.}
    We report the accessible mutual information $I_{\rm acc}(\calS:\calE_1)$ between system and an individual environmental qubit as a function of time, for different values of $p$.
    The model parameters are fixed at $\omega=\gamma=0.1, n=8$. The reported accessible information is rescaled over $S(\rho_\calS)$.}
    \label{fig:accessible_MI_vs_time}
\end{figure}

To study the time-evolution of the system-environment state in this model and its objectivity we will focus on the accessible information $I_{\rm acc}$ and redundancy $\on{Red}$ as quantifiers.
Here $I_{\mathrm{acc}}(\calS:\calF_i)$ is the accessible mutual information between $\calS$ and $\calF_i$, defined as the (classical) mutual information of the joint probability distribution obtained measuring $\rho_{\calS\calF_i}$, maximised over all measurement choices on $\calS$ and $\calF_i$.
This quantity is bounded by the Holevo $\chi$~\cite{holevo_bounds_1973} defined as 
\begin{equation}\scalebox{0.89}{$
\chi(\calS:\calF_i)=\max_{\Pi^\calS}\left\{S\left(\sum_ap_a\rho_{\calF_i|a}\right)-\sum_ap_aS\left(\rho_{\calF_i|a}\right)\right\},
    $}
\end{equation}
where $a$ are the possible measurement outcomes of system measurements $\Pi^\calS$, with $p_a$ the probability of said outcomes and $\rho_{\calF_i|a}$ the corresponding conditioned states of $\calF_i$, with the obvious advantage of needing to maximise over only one set of measurements.
\begin{figure}[t]
    \centering
    \includegraphics[width=0.88\columnwidth]{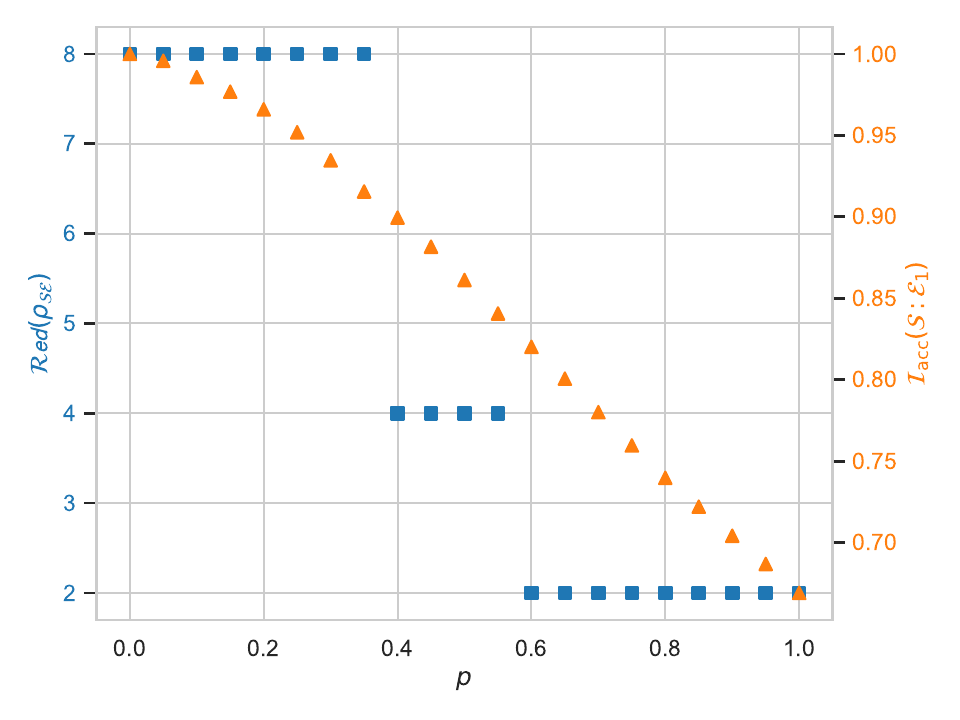}
    \caption{
    \textbf{Maximal redundancy for different values of $p$.}
    Redundancy $\Red(\rho_{\calS\calE})$ and accessible mutual information $I_{\rm acc}(\calS:\calE_1)$ as a function of the non-commutativity parameter $p$, at the time maximising $\Red(\rho_{\calS\calE})$.
    Parameter values are $\gamma=\omega=0.1$, $n=8$ and $\delta=0.9$. Both $\Red(\rho_{\calS\calE})$ and $I_{\rm acc}(\calS:\calE_1)$ are evaluated at the time where they reach their maximum, which depends on $p$ but is the same for both quantities.
    }
    \label{fig:Single}
\end{figure}
On the other hand, the redundancy of a state~\cite{blume-kohout_quantum_2006, riedel2012rise} is a standard way to quantify its objectivity.
An operative way to define redundancy is as the largest number of fractions $\calF_i$ into which we can partition $\calE$, still ensuring each fraction 
has full information about $\calS$. 
More precisely, the redundancy of $\rho_{\calS\calE}$ is defined as~\cite{chisholm2023meaning}
\begin{equation}\scalebox{0.89}{$
    \Red(\rho_{\calS\calE})
    \equiv \max\left\{r:\,
    \calE=\bigotimes_{i=1}^r \calF_i, I_{\rm acc}(\calS:\calF_i)\simeq S(\rho_\calS)\right\},
$}\end{equation}
where $\calF_i$ are in general composed of multiple qubits, in the cases where a single environmental qubit does not hold sufficient 
information about $\calS$.
Here $I_{\rm acc}(\calS:\calF_i)\simeq S(\rho_\calS)$ is to be interpreted here as the condition $I_{\rm acc}(\calS:\calF_i)\ge (1-\delta) S(\rho_\calS)$ for some fixed threshold parameter $\delta>0$.
For readability we will suppress the explicit $\delta$-dependence and only specify its value when needed.
The Hamiltonian we consider couples $\calS$ symmetrically with each $\calF_i$, thus we do not need to distinguish between redundancy and consensus~\cite{chisholm2023meaning, chisholm2024importance}, and safely interpret $\Red(\rho_{\calS\calE})$ as the number of observers simultaneously holding near-complete information about $\calS$.
Although redundancy is a coarse measure of objectivity --- especially for small environments --- its clear operational interpretation makes it one of the most relevant figures of merit here.

\begin{figure}[ht!]
    \centering
    \includegraphics[width=\columnwidth]{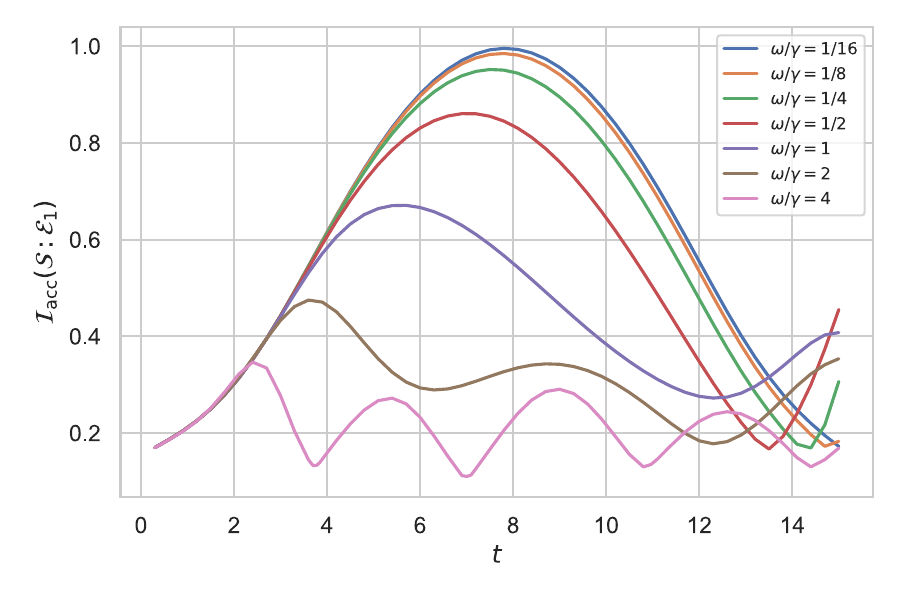}
    \caption{
    \textbf{Accessible information as a function of time.}
    We report the accessible mutual information $I_{\rm acc}(\calS:\calE_1)$ between system and an individual environmental qubit as a function of time, for different values of the $\omega/\gamma$ ratio, fixing $p=1$ and $n=8$. The reported accessible information is rescaled over $S(\rho_\calS)$.
    }
    \label{fig:Rates}
\end{figure}

\Cref{fig:accessible_MI_vs_time} shows the time evolution of the accessible mutual information $I_{\mathrm{acc}}(\calS:\calE_1)$ between $\calS$ and one environment qubit $\calE_1$, for various values of $p$, for $\gamma=\omega=0.1$ and $n=8$.
This clearly shows that for small $p$ one qubit suffices to recover full information about $\mathcal{S}$, but the encoding worsen progressively with increasing $p$, requiring multiple environment qubits to recover the same information.
Furthermore, it is clear from this data that the amount of information that leaked into the environment is strongly time-dependent --- as expected in any unitary dynamics of this kind, due to finite-size effects.

To focus on the maximum degree of objectivity sustained by the dynamic, we then considered the time at which $\Red(\rho_{\calS\calE})$ reaches its first local maximum.
\Cref{fig:Single} shows both $\Red(\rho_{\calS\calE})$ and $I_{\mathrm{acc}}(\calS:\calE_1)$ for different values of $p$, at the time of maximum redundancy. We again use the parameter values $\gamma=\omega=0.1$ and $n=8$, the redundancy is computed using a threshold value of $\delta=0.9$.
These results clearly show how objectivity degrades monotonically with increasing $p$, consistent with the intuition that a changing system state causes information acquired at earlier time to become outdated, thus hindering the emergence of objectivity.
Similar effects of non-commutativity on quantum objectivity were previously observed in settings involving partially inaccessible environments~\cite{ryan2022commutativity}.

\begin{figure}[t]
    \centering
    \includegraphics[width=0.88\columnwidth]{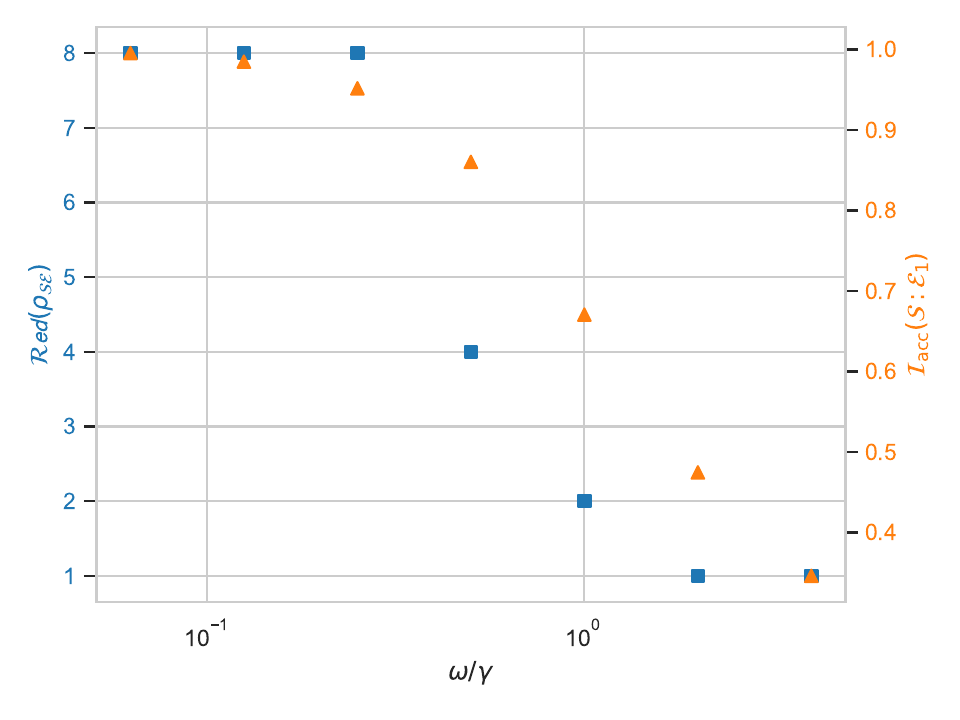}
    \caption{
    \textbf{Maximal redundancy changing model parameters}
    We report $\Red(\rho_{\calS\calE})$ and $I_{\rm acc}(\calS:\calE_1)$ computed at the time of maximal redundancy, as a function of $\omega/\gamma$, fixing $p=1$, $n=8$ and $\delta=0.9$.
    }
    \label{fig:red_vs_parameters}
\end{figure}

To investigate how objectivity depends on the other model parameters, we report in~\cref{fig:Rates} the time evolution of the accessible mutual
information $I_{\mathrm{acc}}(\calS:\calE_1)$, and in in~\cref{fig:red_vs_parameters} the behaviour of $\Red(\rho_{\calS\calE})$ and $I_{\rm acc}(\calS:\calE_1)$ at the time of maximum redundancy, both as a function of $\omega/\gamma$, fixing $p=1$.
The $\omega/\gamma$ ratio characterises the relative speeds of system and interaction Hamiltonians.
For $\gamma\gg\omega$, dephasing outpaces the system's drive: the environment acquires information about $\rho_\calS$ and induces decoherence before said information had a chance to degrade significantly, making the effects of non-commutativity negligible.
On the other hand, when $\gamma\ll\omega$, the system evolves too fast for the environment to have a chance to read it. In this case, non-commutativity becomes highly disruptive and degrades the objectivity.
These results are consistent with the fact that macroscopic environments give rise to objectivity, while it is generally unjustified to assume that $[H_\mathcal{S}, H_\mathcal{I}] = 0$ holds in all such systems.
In the limit $\gamma\gg\omega$ --- expected for large baths --- decoherence happens at much smaller timescales compared to the system's dynamics,  allowing the emergence of objectivity even for non-commuting interactions.
While our results show that non-commutativity is a hindrance to the emergence of objectivity, they also point out that objective states can emerge as a result of non-commutative dynamics, and that in the macroscopic limit the effect of non-commutativity may become negligible.
\paragraph{Pointer states}
\label{sec:pointers}

\begin{figure*}
    \centering
    \includegraphics[trim={0.5cm 0cm 0.4cm 0cm}, clip, width=0.66\columnwidth]{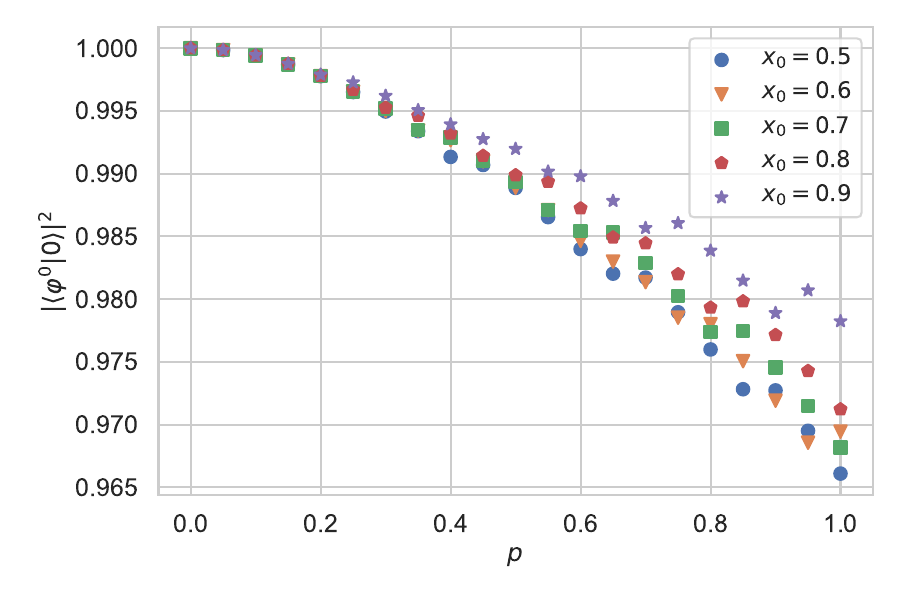}
    \includegraphics[trim={0.5cm 0cm 0.4cm 0cm}, clip, width=0.66\columnwidth]{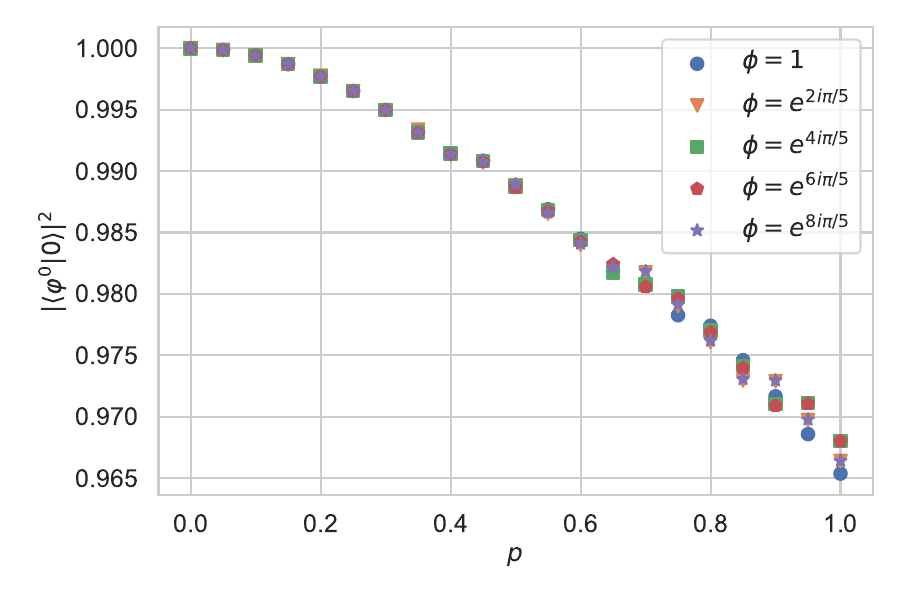}
    \includegraphics[trim={0.5cm 0cm 0.4cm 0cm}, clip, width=0.66\columnwidth]{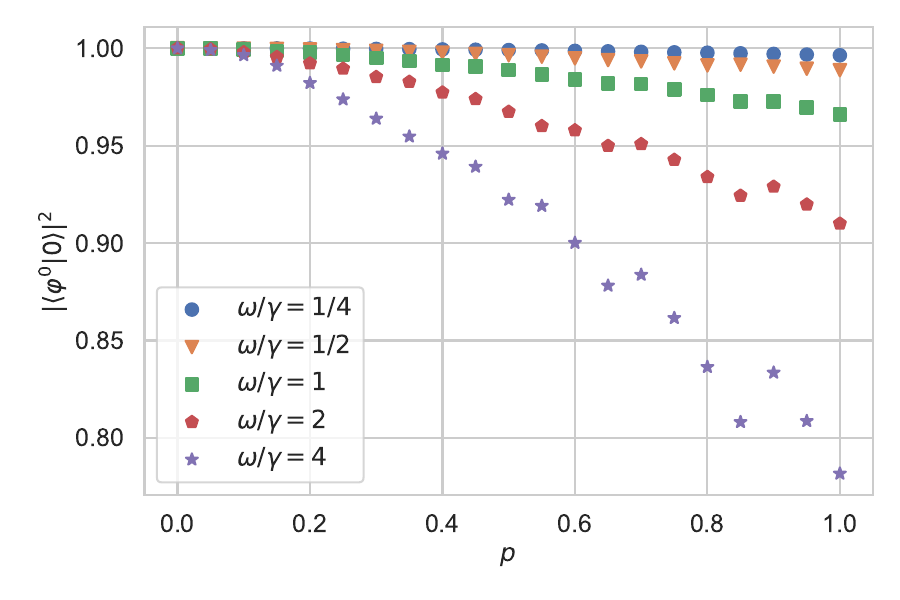}
    \caption{\textbf{Fidelity, with respect to the computational basis, for the pointer states.}
    Fidelity between one of the pointer states emerging from the dynamics $|\varphi^0\rangle$ and $|0\rangle$ as a function of the $p$ parameter. Panel a): the initial state is of the form $\sqrt{x_0}\ket{0}+i\sqrt{1-x_0}\ket{1}$ with $\gamma=\omega=0.1$. Panel b): the initial state is of the form $\frac{1}{\sqrt{2}}\ket{0}+\phi\frac{1}{\sqrt{2}}\ket{1}$ with $\gamma=\omega=0.1$. Panel c): the initial state is $|\circlearrowleft\rangle$ with different degrees of the $\omega/\gamma$ ratios. For all panels $n=6$ and $\delta=0.9$. The plot shows that
    }
    \label{fig:MI}
\end{figure*}

So far, we have shown that objective states may emerge even in the presence of non-commutativity between $H_{\mathcal{S}}$ and $H_\mathcal{I}$.
We now study whether such dynamics also select pointer states, how these should be defined, and what operational meaning can be attached to them.

In their original conception, pointer states are those least affected by environment-induced decoherence~\cite{zurek_decoherence_2003}.
These can be formalised as the $\{|\psi^P_i\rangle\}$ such that, for any environment state $|\psi_{\calE}\rangle$ and all $t$, the evolution maps $|\psi^P_i\rangle\otimes|\psi_{\calE}\rangle\to|\psi_i^P\rangle\otimes|\psi_{\calE}(t)\rangle$.
Thus pointer states remain pure throughout the dynamics, and information about them leaks into the environment.
This definition is incompatible with non-commuting $H_\calS$ and $H_{\mathcal{I}}$~\cite{duruisseau2023pointer, doucet2024classifying, brasil2015understanding}, as it is not possible to find system states that are simultaneously eigenstates of both $H_\calS$ and $H_\calI$.

Another defining feature of pointer states is that they are the states the environment has the most information about; quoting~\cite{zurek1982environment}: ``It is the propagation of the correlations with the pointer basis states which is ultimately responsible for the choice of the pointer observable'' --- they are the system states that would emerge as objective as a result of measurements on the environment~\cite{dalvit2001unconditional, zurek_quantum_2009}.
Pointer states were thus originally introduced to refer to the states corresponding to the possible display positions of a measurement apparatus~\cite{zurek_pointer_1981}.

These definitions are equivalent when $[H_\calS,H_\calI]=0$, where we can identify $\ket*{\psi_i^P}$ as the eigenstates of $H_\calS$ and observe that these states both remain pure and are the ones that the environment gets the most information about.
However, in more general cases with $[H_\calS,H_\calI]\neq0$, there might not be a system basis that is fixed by the dynamic, despite information about the system still being accessible from the environment.
In this sense, the requirement that pointer states ``suffer the least decoherence'' may be excessively restrictive, and does not allow for their identification in contexts where states possessing key pointer states features exist.

The SBS formalism is particularly well-suited to describe states where the environment holds redundant information the system state.
From this perspective, a system-environment state is objective when it can be written as~\cite{korbicz_objectivity_2014}
\begin{equation}
\rho_{\mathrm{SBS}}=\sum_ip_i|\psi_i\rangle\langle\psi_i|\bigotimes^{N}_jR_i^j,
    \label{sbs_form}
\end{equation}
with $\{|\psi_i\rangle\}$ a set of orthogonal system states, and $R_i^j$ states of environmental fractions such that $R_i^jR_{i'}^j=0$ when $i\neq i'$.
The $\{|\psi_i\rangle\}$ states define the basis in which we need to measure the system in order to maximise the mutual information between the system and the environmental fraction, and are therefore the states that the environment has the most information about.

Thus, the SBS formalism provides a less stringent definition of pointer states:
so long as the system-environment state can be cast in SBS form, pointer states are always well defined. In the case where commutativity is restored, SBS-pointer states will also be the ones most resilient to decoherence.
It is worth noting that in the non-commuting scenario, the SBS-pointer basis can no longer be determined \emph{a priori} from the Hamiltonian alone, due to the resulting complex dynamics, rather it must be obtained by casting the joint system-environment state in SBS form.

With this alternative definition of pointer states, valid also for non-commuting evolutions, we can now analyse the pointer states that emerge as a result of the system and environment evolving under the Hamiltonian in Eq.~\ref{eq1}.
We consider three different scenarios.
In the first one, the initial state of the system is of the form $\sqrt{x_0}\ket{0}+i\sqrt{1-x_0}\ket{1}$, with $x_0$ ranging from $x_0=0.5$ to $x_0=0.9$. In the second scenario the initial state of the system lays in the equatorial line of the Bloch sphere, and is of the form $\frac{1}{\sqrt{2}}\ket{0}+\phi\frac{1}{\sqrt{2}}\ket{1}$ with $\phi$ a complex number of unit norm. In both first and second scenarios, $\gamma=\omega=0.1$. Finally in the third case the initial state of the system is $\ket{\circlearrowleft}$, but we have different ratios between the interaction strength $\gamma$ and the system frequency $\omega$. In all scenarios the environment consists in $n=6$ qubits all initially in the $|0\rangle$ state.

We show in~\cref{fig:MI} the fidelity between one of the two pointer states emerging from the dynamics, $\ket{\varphi^0}$, and $\ket{0}$, for different values of the non-commutativity parameter $p$. The basis $\{\ket{0}, \ket{1}\}$, being the eigenstates of the interaction Hamiltonian $H_\mathcal{I}$, would be the pointer states emerging from the dynamics in the presence of commutativity, however, when $H_\calS$ and $H_{\mathcal{I}}$ don't commute the system evolves in an objective state but the resulting pointer states are different.

\Cref{fig:MI}~(a) shows that the resulting pointer basis does not only depend on $H_\calS$ and $H_{\mathcal{I}}$ but also on the initial state of the system, further showing how pointer states cannot be determined \textit{a priori} from the Hamiltonian in the presence of non-commutativity. This dependence on the initial state seem however minimised whenever the system is initially in an equal weight superposition of $|0\rangle$ and $|1\rangle$, i.e. one of the states that suffers the most decoherence from the environment, as shown in~\cref{fig:MI}~(b).
Finally,~\cref{fig:MI}~(c) shows how different driving strengths have a large influence in the resulting pointer states, with pointer states being minimally altered by non-commutativity for $\gamma\gg\omega$, while they change significantly for $\omega\gg\gamma$.
This is expected for the same reasoning outlined in the previous section.

If we interpret the interaction with the environment as a measurement operation carried out by an apparatus, what we can see here is that in the presence of non-commutativity the apparatus is measuring something different than what initially intended, in that the apparatus was designed to measure in the $\{|0\rangle ,|1\rangle\}$ basis and is instead measuring in a different one, making the measurement operation imperfect.
This however cannot be avoided if we want to measure a system in a basis that does not commute with the system's own Hamiltonian, making this one of the many systematic errors that real measurements are subject to.
What we show here is that the fidelity between the intended measurement and the effective one can be high in a wide range of parameters.

\paragraph{Conclusions}

In this work we analysed the emergence of objectivity in the case where the system Hamiltonian and the system-environment interaction Hamiltonian do not commute, something that would naturally happen in many realistic situations. 
We showed that non-commutativity is generally a hindrance to the emergence of objectivity, notwithstanding objective states may emerge as a result of such non-commuting evolutions.

The studied model provided an interesting platform to further explore the concept of pointer states, which are understood as the ones whose information has proliferated into the environment, as well as the states most resilient to environmental decoherence. We showed that if we relax the condition of resilience from decoherence, and simply require proliferation of information, pointer states are always well defined whenever the system is objective according to quantum Darwinism, and pointer states can be identified within the SBS framework.

Our work is a further step into exploring the emergence of objectivity and the quantum-to-classical transition for non-trivial, non-ideal scenarios, and offers a new perspective on the meaning of pointer states.
The analysis of generic properties of these SBS-pointer states remains an interesting research avenue for future work.

\acknowledgments
DAC acknowledges support from  the ``Italian National Quantum Science and Technology Institute (NQSTI)" (PE0000023) - SPOKE 2 through project ASpEQCt. GMP acknowledges support by MUR under PRIN Project No. 2022FEXLYB. Quantum Reservoir Computing (QuReCo).

\bibliography{Ref}
\end{document}